\documentclass[preprint,12pt]{elsarticle}

\usepackage{amssymb}
\usepackage{amsmath}

\journal{Computational Materials Science}

\begin{document}

\begin{frontmatter}



\title{Exotic 4$f$ correlated electronic states of Ferromagnetic Kondo Lattice Compounds \textit{Re}Rh$_6$Ge$_4$ (\textit{Re}=Ce, Ho, Er, Tm)}


\author[label1,label2]{Caiqun Wang} 
\author[label3]{Yu Gao} 
\author[label2]{Jun Jiang} 
\author[label3]{Qiaoni Chen$^\dagger$} 
\ead{qiaoni@bnu.edu.cn}
\author[label4]{Haiyan Lu$^\ddagger$} 
\ead{hyluphys@163.com}
\author[label1]{Ping Qian$^\ast$} 
\ead{qianping@ustb.edu.cn}
\affiliation[label1]{organization={Department of Physics, University of Science and Technology Beijing},
            city={Beijing},
            postcode={100083}, 
            country={China}}

\affiliation[label2]{organization={Beijing Computing Center Co. Ltd},
            city={Beijing},
            postcode={100094}, 
            country={China}}

\affiliation[label3]{organization={Department of Physics, Beijing Normal University},
            city={Beijing},
            postcode={100875}, 
            country={China}}

\affiliation[label4]{organization={Institute of Materials, China Academy of Engineering Physics},
	    city={Mianyang, Sichuan},
            postcode={621907}, 
            country={China}}

\begin{abstract}
CeRh$_6$Ge$_4$ stands out as the first stoichiometric metallic compound with a ferromagnetic quantum critical point, thereby garnering significant attention. Ferromagnetic Kondo lattice compounds $Re$Rh$_6$Ge$_4$ ($Re$=Ce, Ho, Er, Tm) have been systematically investigated with density functional theory incorporating Coulomb interaction $U$ and spin-orbital coupling. We determined the magnetic easy axis of CeRh$_6$Ge$_4$ is within the $ab$ plane, which is in agreement with previous magnetization measurements conducted under external magnetic field and $\mu$SR experiments. We also predicted the magnetic easy axes for the other three compounds. For TmRh$_6$Ge$_4$, the magnetic easy axis aligns along the $c$ axis, thus preserving the $C_3$ rotational symmetry of the $c$ axis. Especially, there are triply degenerate nodal points along the $\Gamma-A$ direction in the band structure including spin-orbital coupling. A possible localized to itinerant crossover is revealed as 4$f$ electrons increase from CeRh$_6$Ge$_4$ to TmRh$_6$Ge$_4$. Specifically, the 4$f$ electrons of TmRh$_6$Ge$_4$ contribute to the formation of a large Fermi surface, indicating their participation in the conduction process. Conversely, the 4$f$ electrons in HoRh$_6$Ge$_4$, ErRh$_6$Ge$_4$ and CeRh$_6$Ge$_4$ remain localized, which result in smaller Fermi surfaces for these compounds. These theoretical investigations on electronic structure and magnetic properties shed deep insight into the unique nature of 4$f$ electrons, providing critical predictions for subsequent experimental studies.
\end{abstract}




\end{frontmatter}



\section{Introduction}\label{sec1:Introduction}
Quantum criticality in heavy fermion materials has been a subject of ongoing experimental and theoretical research in the past twenty years. Heavy fermion materials usually contain lanthanides or actides elements which involve partially filled $f$ orbitals, so they are the typical strongly correlated electron systems. Abundant exotic phenomena have been intensively studied, such as unconventional superconductivity \cite{Ran2019_Science365-684,Jiao2020_Nature579-523--527,LiYu2021_ActaPhys.Sin.70-106-136,jiaolin:586}, non-Fermi liquid behaviors \cite{legros:cea-02086419,PhysRevLett.85.626,Daou_2008,RevModPhys.73.797,PhysRevLett.72.3262,PhysRevLett.85.626}, and exotic quantum critical phenomena \cite{Si2010_Science329-1161--1166,Liu2023_PNAS120-e2300903120,Custers2003_Nature424-524--527}. In some heavy fermion materials quantum criticality is largely driven by the competition between RKKY interaction and Kondo effect, and this mechanism falls in Landau's paradigm. However there is a large class of heavy fermion materials, on which the quantum critical point (QCP) is accompannied by the sudden breakdown of Kondo correlation. In the scenario of Kondo breakdown QCP, a jump occurs in the Fermi volume from large to small at the critical point. The Hertz-Millis-Moriya theory \cite{Hertz1976_PRB14-1165--1184,Millis1993_PRB48-7183--7196,Moriya1985} made foundation for the quantum critical phenomena of the itinerant fermions. In the ferromagnetic metallic materials, the ferromagnetic phase is either enter into other ordered phase first, or enter into the paramagentic phase through a first order phase transition \cite{Brando2016_RMP88-025006,Belitz1999_PRL82-4707--4710,PhysRevB.91.035130,PhysRevLett.105.217201,PhysRevLett.93.256404}. In the framework of Hertz-Millis-Moriya theory, two groups predicted that the ferromagnetic QCP is not stable in itinerant syste\cite{Belitz1999_PRL82-4707--4710,Chubukov2004_PRL92-147003}. However recent experiments on CeRh$_{6}$Ge$_{4}$ show clear evidence of ferromagnetic QCP under high pressure \cite{Shen_2020}.

The germanide compound CeRh$_{6}$Ge$_{4}$ is a ferromagnetic metal, and it was synthesized from the elements by Bisbuth fluxes\cite{Vosswinkel2012_ZfNatB67-1241}. Resistivity measurements under high pressure \cite{Kotegawa2019_JPSJ88-093702} first suggest there exists ferromagnetic QCP in CeRh$_{6}Ge_{4}$. Later detailed thermodynamic and transport measurements confirm the existence of the ferromagnetic QCP in CeRh$_6$Ge$_4$ under high pressure \cite{Shen_2020}. Then by silicon doping people also find ferromagentic QCP \cite{Zhang2022_PRB106-054409} under chemical pressure. Although recent theoretical work suggests that if the material is noncentrosymmetric with strong spin-orbit coupling, is quasi-one-dimensional, or is sufficiently dirty, i.e., has a short electronic mean-free path, ferromagentic QCP is possible to appear in itinerant electrons system \cite{Kirkpatrick2020_PRL124-147201}. The nature of the ferromagentic QCP on CeRh$_{6}Ge_{4}$ is still under debate. Whether it belong to the Kondo breakdown criticality \cite{Komijani2018_PRL120-157206,Shen_2020} , or there are two sequential effects near QCP as suggested by recent thermopower measurement\cite{Thomas2024_PRB109-L121105}. Recent numerical research indicate the anisotropy could lead to ferromagentci QCP \cite{Chen2022_PRB106-075114,Wang2022_SCPMA65-257211}, meanwhile both angle-resolved photoemission spectroscopy (ARPES) \cite{Wu_2021} and ultrafast \cite{Pei2021_PRB103-L180409} optical experiments suggest the anisotropic hybridization of CeRh$_{6}$Ge$_{4}$.       

Besides CeRh$_{6}$Ge$_{4}$ other isostructure compounds $Re$Rh$_{6}$Ge$_{4}$, by substituting of $Ce$ element into other lanthanides elements have been synthesized \cite{Vosswinkel2013_Zfauac639-2623}. At the same time they all have the LiCo$_6$P$_4$ type structure \cite{buschmann1991darstellung}, and display rich exotic properties. Transport meansuments indicate that along different direction the charge carriers of LaRh$_6$Ge$_4$ are with different type \cite{Luo2023_PRB108-195146}. La-doped CeRh$_6$Ge$_4$ display the crossover from coherent Kondo lattice behaviors to Kondo impurity regime \cite{Xu2021_CPL38-087101}. DFT calculations imply there are non-trivial topological properties in the band structure of YRh$_{6}$Ge$_{4}$, LaRh$_{6}$Ge$_{4}$ and LuRh$_{6}$Ge$_{4}$ \cite{Guo2018_PRB98-045134}. Though YRh$_{6}$Ge$_{4}$, LaRh$_{6}$Ge$_{4}$ and LuRh$_{6}$Ge$_{4}$ are non-magnetic, several other $Re$Rh$_{6}$Ge$_{4}$ have local mangnetic moments. Among them HoRh$_{6}$Ge$_{4}$, ErRh$_{6}$Ge$_{4}$ and TmRh$_{6}$Ge$_{4}$ all have ferromagnetic ground state, same as CeRh$_{6}$Ge$_{4}$. In this paper we run the DFT calculations on the these four ferromagnetic germanide compounds. We first calculate the total ground state energy along different magnetic axes, and predict the easy axis of these compounds. Then we analysis the band structure, and find the nontrivial topological points in the band structure. In the end we calculate the DOS and Fermi surface, and find the evolution of the Fermi surface as the 4$f$ electrons increase. 
       	
\section{COMPUTATIONAL DETAILS}\label{sec2:Computional_Detail}
First-principles electronic structure calculations were performed using the fully potential (linear) augmented plane wave (LAPW)$+$local orbital (lo) method implemented in the WIEN2k package \cite{wien2k}. Exchange correlation functionals were performed using Perdew-Burke-Ernzerhof (PBE) type of generalized gradient approximation (GGA) \cite{PhysRevLett.77.3865}. A 7$\times$7$\times$11 k-points mesh is taken for the full Brillouin zone (BZ) sampling, and the maximum cut of the wave vector was adjusted to R$_{mt}$K$_{max}$ = 8.5. Murnaghan equation of state is utilized to obtain the zero temperature equilibrium structures. Once the equilibrium structures were obtained, the self-consistent calculations with an energy convergence value of $10^{-6}$ Ry and a charge convergence value of $10^{-4}$ were performed. Since the lathanides elements are heavy, the spin-orbital coupling (SOC) are considered. Due to the partially filled $4f$ orbital on the rare earth ions, the Hubbard U are also included in the calculations. So the SOC and Hubbard $U$ are both included on the rare earth ions. Ferromagnetic calculations are performed for all of the four compounds, as they all have ferromagnetic ground state. However a knotty issue in the DFT calculations of the system with $4f$ electrons is that the energy curved surface can be complex. Sometimes the calculation are difficult to converge, and sometimes they get stuck in local minima \cite{dorado2013advances}. To overcome these problems, we carefully chose the initial electronic configuration when searching for the ground state along different magnetic axes and the results are reliable.

\section{RESULTS AND ANALYSIS}\label{sec3:Result-And-Analysis}
\begin{figure}[htbp]
		\includegraphics[width=0.98\textwidth]{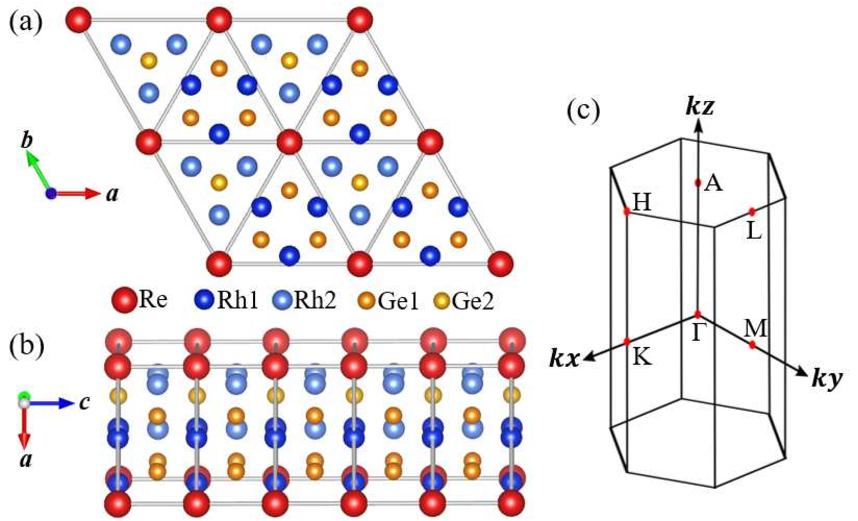}
		\caption{\label{fig:struct}(a) Top view and (b) perspective view along the c-axis of the crystal structure of $ReRh_6Ge_4$ ($Re=Ce,Ho,Er,Tm$). The red sites are rare-earth ions, and they form triangular lattice in the $ab$ plane. The blue sites are $Rh$ ions, and the yellow sites are $Ge$ ions. (c) The first Brillioun zone of $ReRh_6Ge_4$, and the red points are the high symmetry points.}
		\label{struct}
	\end{figure}

The lattice structure of the Kondo lattice compounds $ReRh_{6}Ge_{4}$ ($Re=Ce,Ho,Er,Tm$) belong to the hexagonal crystal system. The rare-earth ions forms triangular lattice in the $ab$ plane, and the whole lattice are composed of a stack of triangular lattices along $c$ axis as displayed in Fig. ~\ref{fig:struct}. The lattice constant along the $c$ axis is much shorter than it along the $a$ axis or $b$ axis as shown in Table ~\ref{tab:table1}. The $ReRh_{6}Ge_{4}$ lattice belong to the non-centrosymmetric space group $P\bar{6}m2$, and is without space-inversion symmetry. As there are two kinds of inequivalent $Rh$ irons and $Ge$ irons, marked with light (dark) blue sites and light (dark) yellows sites
in Fig. ~\ref{fig:struct}, the lattice is non-centrosymmetric. The first Brillioun zone is a hexagonal prism as displayed in Fig. ~\ref{fig:struct} (c),  the $k_z$ direction is just along the $c$ axis, the $k_x$ direction is along the $a$ axis, and the $y$ direction is perpendicular to the $x$ direction. The red points are the high symmetry points in the Brillioun zone. The distance of $\Gamma-A$ is longer since the $c$ axis lattice constant is shorter. 

The lattice constants of these Kondo lattice compounds ReRh$_6$Ge$_4$ are optimized by the Murnaghan equation of state \cite{murnaghan1944compressibility} in the paramagnetic phase. The optimizition start from the data of X-ray diffraction experiments on the powder samples \cite{Vosswinkel2013_Zfauac639-2623}, and our results are displayed in Table ~\ref{tab:table1}. As the atomic number of rare-earth elements increases from CeRh$_6$Ge$_4$ to TmRh$_6$Ge$_4$,  the lattice constants along $c$ axis become smaller, while the lattice constants along $a$ axis tend to become bigger. The ratio of $a/c$ becomes smaller, except from ErRh$_6$Ge$_4$ to TmRh$_6$Ge$_4$. The volume of the unit cell becomes smaller as a result of the radius of the rare-earth ions decease.

	\begin{table}
		\centering
		\caption{\label{tab:table1}%
			Stable lattice parameters of the rare earth germanides $ReRh_6Ge_4$ ($Re$ = Ce, Ho, Er, Tm). }
			\begin{tabular}{ccccc}
				\hline
				Compound& $a$ /\r A  &$c$ /\r A &$a/c$ &Volume / \r A$^3$\\ \hline
				CeRh$_6$Ge$_4$ & 7.201 & 3.881 & 1.855 & 174.282\\
				HoRh$_6$Ge$_4$ & 7.236 & 3.840 & 1.884 & 174.124\\
				ErRh$_6$Ge$_4$ & 7.239 & 3.836 & 1.887 & 174.087\\
				TmRh$_6$Ge$_4$ & 7.238 & 3.836 & 1.886 & 174.039\\\hline
			\end{tabular}
	\end{table}
		
	\subsection{Ferromagnetic Ground State and Magnetic Easy Axis}\label{subsec1:Ferromagnetic}
Magnetic measurements indicates several rare-earth germanides are with local magnetic moments \cite{Vosswinkel2013_Zfauac639-2623}, such as GdRh$_6$Ge$_4$, TbRh$_6$Ge$_4$, DyRh$_6$Ge$_4$ and YbRh$_6$Ge$_4$. Among them CeRh$_6$Ge$_4$ and the other three compounds we studied have ferromagnetic ground state. CeRh$_6$Ge$_4$ has been studied intensively, and its magnetic easy axis is within the $ab$ plane according to the magnetization measurements \cite{Shen_2020}. The magnetic Bragg peaks is not resolved in neutron diffraction of powder sample, but the coherent oscillations are observed in zero-filed $\mu SR$ measurements \cite{Shu2021_PRB104-L140411}. The $\mu SR$ measurements suggest the magnetic easy axis is along the a axis. In the following we implemented ferromagnetic DFT calculations to look for the magnetic easy axis. The spin-orbital coupling is included on $Ce$ as it's a heavy element, and Hubbard $U$ is fixed to $6$ $eV$ as previous work \cite{Wu_2021,WANG20211389}. Since lanthanides elements have unfilled $4f$ shells, the magnetism on these rare-earth germanides is mainly contributed by the lanthanides elements.     

	\begin{figure}[htbp]
		\centering
		\includegraphics[width=0.43\textwidth]{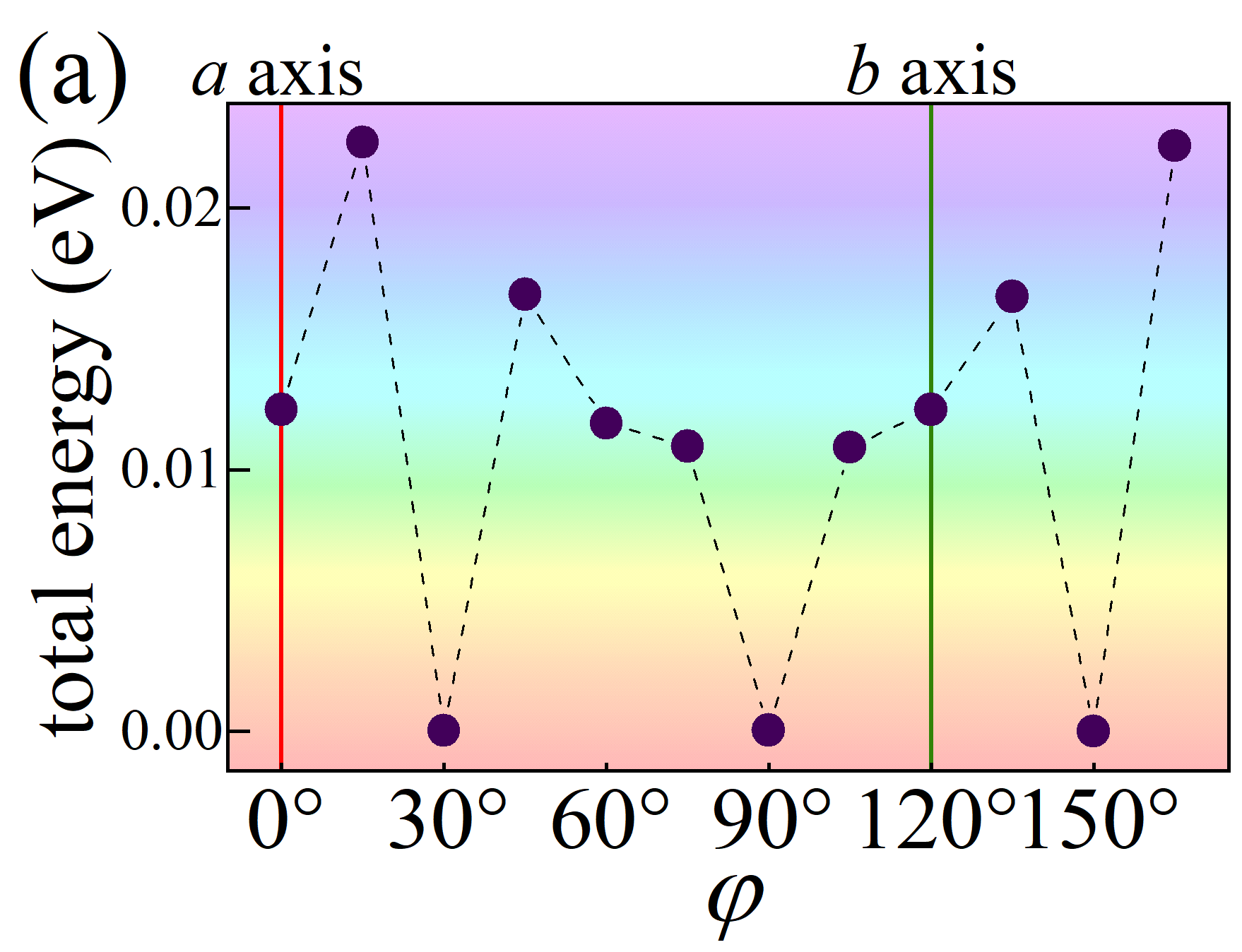}
		\includegraphics[width=0.43\textwidth]{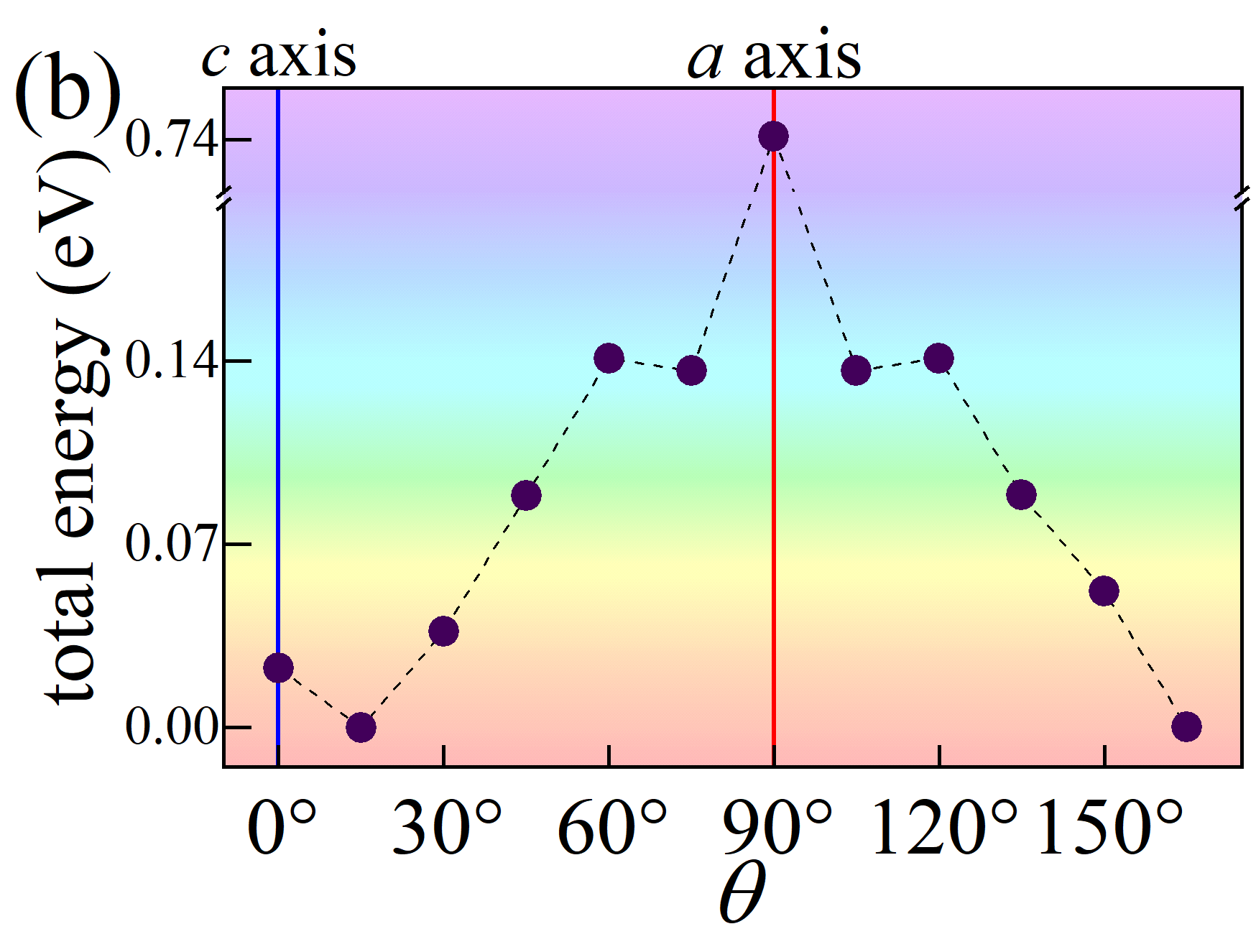}
		\includegraphics[width=0.43\textwidth]{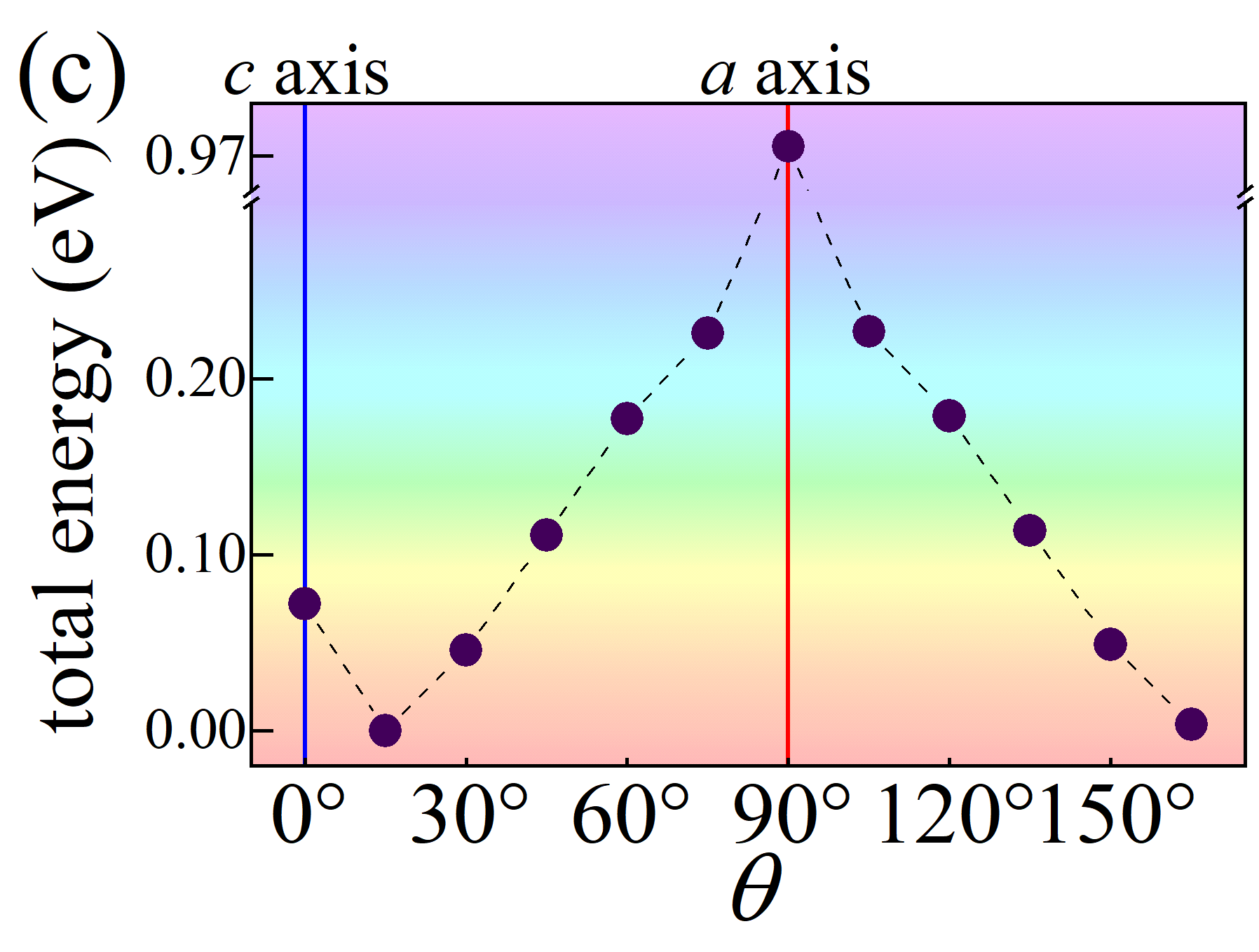}
		\includegraphics[width=0.43\textwidth]{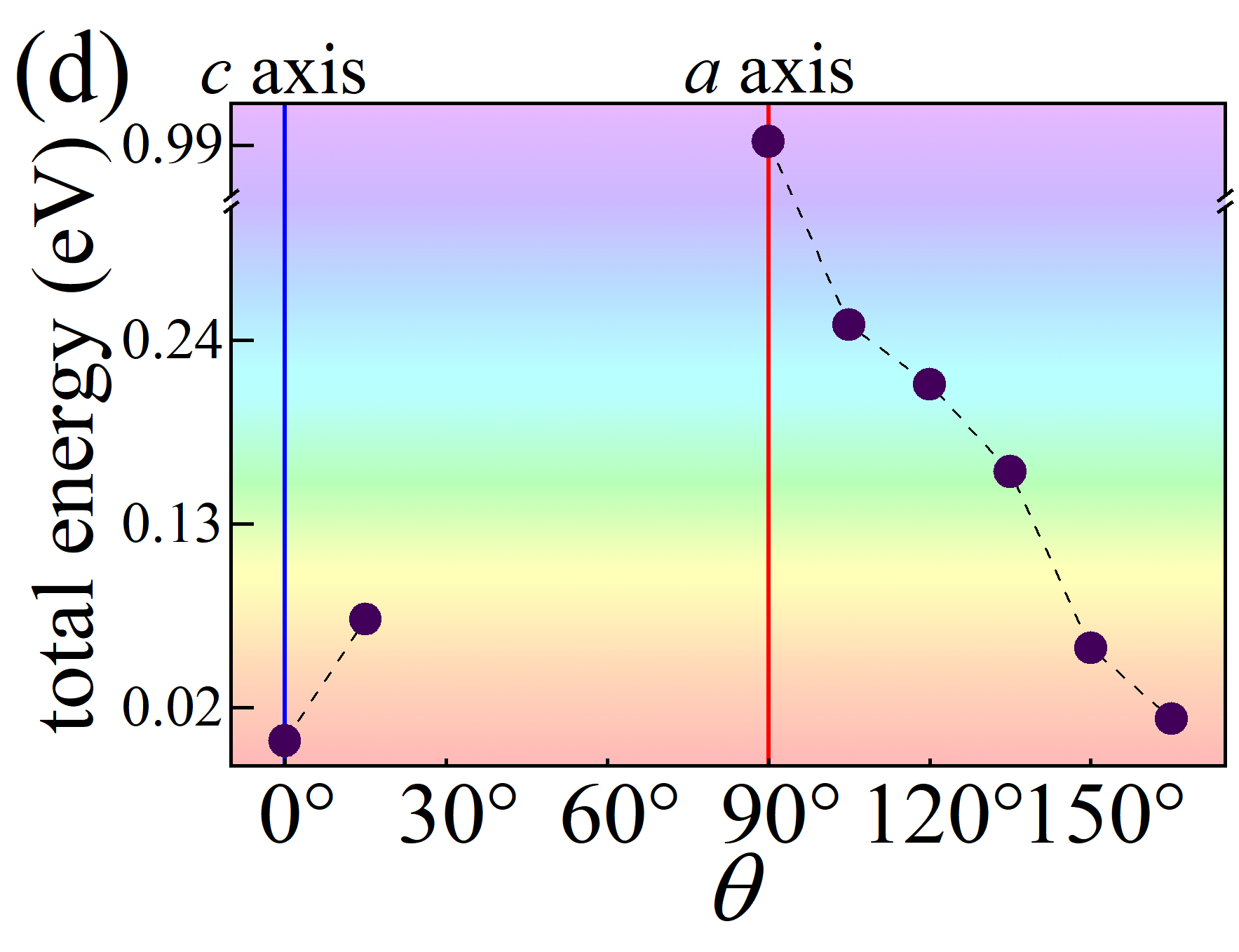}
		\caption{\label{fig:ENE} The total ground state energy along different magnetic axis. (a) The ground state energy of CeRh$_6$Ge$_4$ when the magnetic axis is within the $ab$ plane. $\varphi$ is the angle with the $a$ axis, so the red line and green line represent the $a$ axis and the $b$ axis respectively. (b)-(d) The ground state energy of HoRh$_6$Ge$_4$, ErRh$_6$Ge$_4$ and TmRh$_6$Ge$_4$ when the magnetic axis is within the $ac$ plane. $\theta$ is the angle with the $c$ axis, so the blue line and the red line correspond to the $c$ axis and the $a$ axis, respectively.}
	\end{figure} 
	
We first calculate the ground state energy when the magnetic axis is along the lattice vector as shown in Table \ref{tab:table2}. The total ground state energy is $-1250481.63791$ $eV$ when the magnetic axis is along the $c$ axis, while the total ground state energy is $-1250481.70971$ $eV$ and $-1250481.70971$ $eV$ along the $a$ axis and $b$ axis. When the magnetic axis is along the $c$ axis the ground state energy is much higher, it indicates our calculations are in accordance with the experiments. The ground state the $ab$ plane. The $a$ axis is chosen as the $x$ axis, thus the $a$ axis correspond to $\varphi=0^{\circ}$ and the $b$ axis correspond to $\varphi=120^{\circ}$. The ground state energy of CeRh$_6$Ge$_4$ is displayed in Fig. \ref{fig:ENE} (a), when the magnetic axis is within the $ab$ plane. When $\varphi=30^{\circ}$, $\varphi=90^{\circ}$ and $\varphi=150^{\circ}$, the ground state energy is the lowest. In Fig. \ref{fig:ENE} (a) the ground state energy subtract the ground state energy when $\varphi=30^{\circ}$, and the value is $-1250481.72198$ $eV$. Thus our calculations imply the magnetic easy axis of CeRh$_6$Ge$_4$ is along $\varphi=30^{\circ}$, $\varphi=90^{\circ}$ and $\varphi=150^{\circ}$. 

	\begin{table}
		\centering
		\caption{\label{tab:table2}%
			Ferromagnetic ground state energy of the rare earth germanides ReRh$_6$Ge$_4$ ($Re$ = Ce, Ho, Er, Tm) along different axis. }
			\begin{tabular}{ccccc}
				\hline
                Compound & $a$ axis ($eV$)  & $b$ axis ($eV$)   & $c$ axis ($eV$)    \\ \hline   
				CeRh$_6$Ge$_4$ & -1250481.70971 & -1250481.70971 & -1250481.63791 \\
				HoRh$_6$Ge$_4$ & -1352842.01992 & -1352842.01993 & -1352843.17190 \\
				ErRh$_6$Ge$_4$ & -1365658.67112 & -1365658.67112 & -1365659.57456 \\
				TmRh$_6$Ge$_4$ & -1378779.41052 & -1378779.41052 & -1378780.40282 \\\hline
			\end{tabular}
	\end{table}

The magnetization and susceptibility measurements \cite{Vosswinkel2013_Zfauac639-2623} on HoRh$_6$Ge$_4$, ErRh$_6$Ge$_4$ and TmRh$_6$Ge$_4$ indicate there are local magnetic moments on these compounds. The Curie temperature of HoRh$_6$Ge$_4$ and ErRh$_6$Ge$_4$ are all above zero, it suggests ferromagnetism on these two compounds. Both CeRh$_6$Ge$_4$ and TmRh$_6$Ge$_4$ have smaller local magnetic moment, so from the higher temperature magnetic measurements the Curie temperature is below zero. However recent experiments clarify the Curie temperature of CeRh$_6$Ge$_4$ is near $2.5K$. The local magnetic moment of TmRh$_6$Ge$_4$ is larger than CeRh$_6$Ge$_4$, thus the Curie temperature of TmRh$_6$Ge$_4$ should be higher than $2.5K$. Although the magnetic measurements are rare on HoRh$_6$Ge$_4$, ErRh$_6$Ge$_4$ and TmRh$_6$Ge$_4$, they all behave ferromagnetism. We have run the ferromagnetic DFT calculations on these three compounds. For TmRh$_6$Ge$_4$, the total ground state energy is $-1378780.40282$ $eV$ when the magnetic axis is along the $c$ axis, smaller than $-1378779.41052$ $eV$ and $-1378779.41052$ $eV$ along the $a$ axis and $b$ axis. While for HoRh$_6$Ge$_4$ and ErRh$_6$Ge$_4$, the total ground state energy along the $c$ axis is also smaller than that along the $a$ axis or $b$ axis as shown in Table \ref{tab:table2}. Thus be different from CeRh$_6$Ge$_4$, the magnetic easy axis of the other three compounds is no longer within the $ab$ plane. So we scan the ground state energy within the $ac$ plane, and the results are shown in Fig. \ref{fig:ENE}. The $c$ axis is chosen as the $z$ axis, thus the $c$ axis correspond to $\theta=0^{\circ}$ and the $a$ axis correspond to $\theta=90^{\circ}$, marked with blue line and red line respectively. On HoRh$_6$Ge$_4$ and ErRh$_6$Ge$_4$ the ground state energy is lowest when $\theta=15^{\circ}$ and $\theta=165^{\circ}$ , while on TmRh$_6$Ge$_4$ the ground state energy is lowest when $\theta=0^{\circ}$. When the magnetic axis is along $\theta=15^{\circ}$ the ground state energy of HoRh$_6$Ge$_4$ and ErRh$_6$Ge$_4$ is $-1352843.19446$ $eV$ and $-1365659.64667$ $eV$ respectively, as are subtracted in Fig. \ref{fig:ENE}(b) and \ref{fig:ENE}(c). The ground state energy of TmRh$_6$Ge$_4$ is $-1378780.40282$, when the magnetic axis is along the $c$ axis.

	\subsection{Band structure and potential topological properties}\label{subsec2:Bandstructure}
\begin{figure}[htbp]
		\centering
		\includegraphics[width=0.68\textwidth]{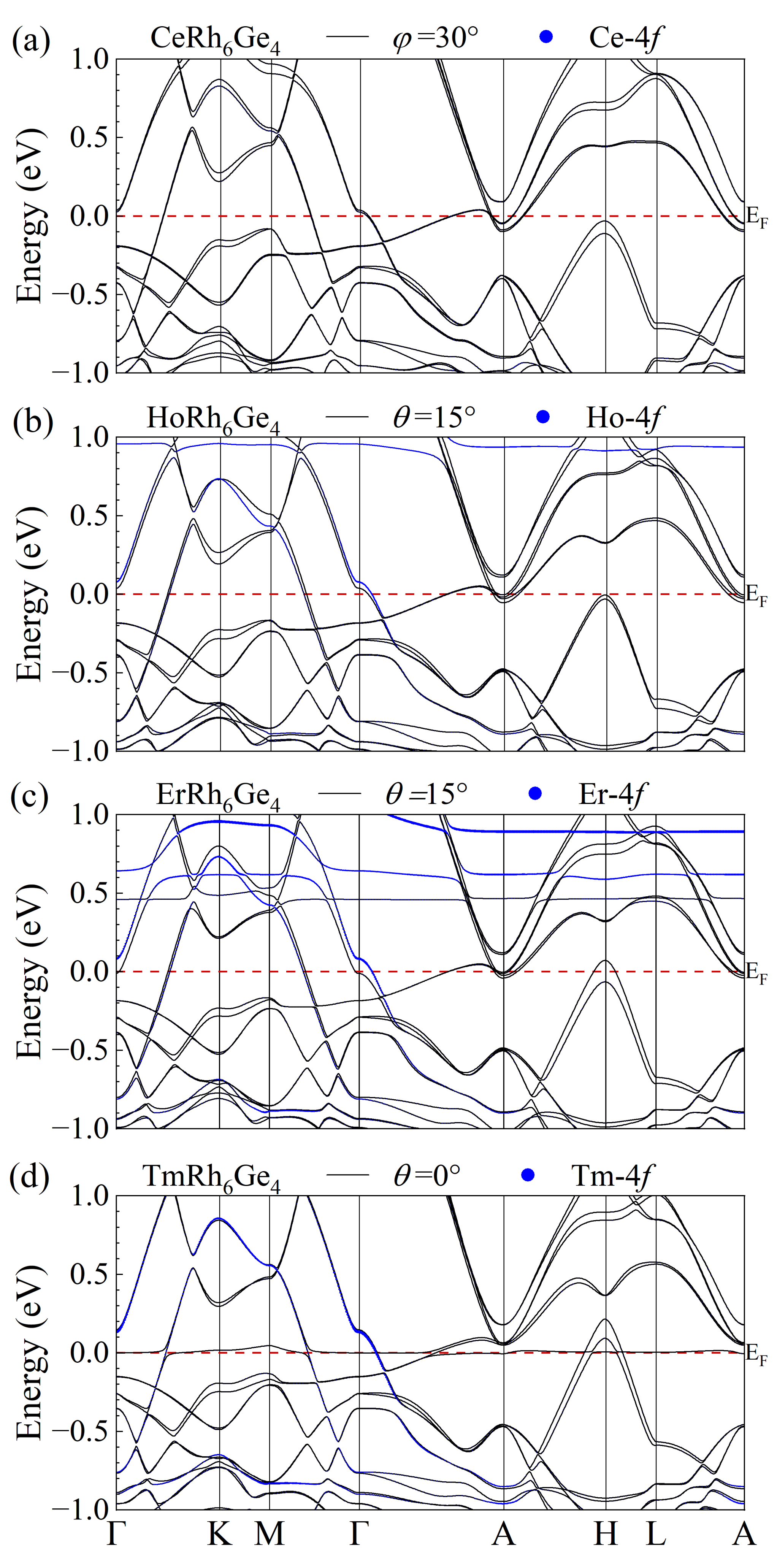}\\
		\caption{\label{fig:band} Band structure of (a) CeRh$_6$Ge$_4$, (b) HoRh$_6$Ge$_4$, (c) ErRh$_6$Ge$_4$ and (d) TmRh$_6$Ge$_4$ along high symmetry paths in the BZ. The calculations are carried in the ferroamgnetic phase with the SOC, and by fixed $U=6$ $eV$. These compounds are all metallic, with bands crossing the Fermi surface (red dashed lines). The blue lines are the $Re$-$4f$ orbital projected band structure. }		
	\end{figure}

Previous DFT simulations on CeRh$_6$Ge$_4$ focus on the paramagnetic phase. We first perform the paramagnetic calculations and compare them with previous simulations, the band structure of our full potential full electron DFT calculations is consistent with these previous calculations. As the ferromagnetism play an important role in CeRh$_6$Ge$_4$, here we mainly concentrate on the ferromagnetic calculations. The ground energy along different axis indicate the easy axis is along $\varphi=30^{\circ}$, so we choose $\varphi=30^{\circ}$ as the magnetic axis. The spin-orbital coupling can not be ignored on $Ce$, and neither the Hubbard $U$. If $U$ is small the bands mainly contributed by $4f$ orbitals are quite close to the Fermi energy, and this is contraditionary with the quantum oscillation experiment \cite{WANG20211389}. In this paper we fix $U=6$ $eV$ following the previous simulations \cite{Wu_2021}. The band structure of CeRh$_6$Ge$_4$ are displayed in Fig. \ref{fig:band} (a) Several bands cross the Fermi energy, so CeRh$_6$Ge$_4$ is metallic as revealed by experiments. In order to see the contribution of the $4f$ electrons, the band structure is also projected to $4f$ orbitals. The $4f$ orbitals of the $Ce^{3+}$ ions barely contribute to the bands near Fermi energy. As the electron configuration of Ce$^{3+}$ is $4f^1$, the band constitute by $4f$ orbitals would be much lower than the Fermi level.           	
	
The band structure of HoRh$_6$Ge$_4$ and ErRh$_6$Ge$_4$ are displayed in Fig. \ref{fig:band} (b) and \ref{fig:band} (c). According to the band structure, both HoRh$_6$Ge$_4$ and ErRh$_6$Ge$_4$ are metals since several bands cross the Fermi level. The electron configuration of Ho$^{3+}$ and Er$^{3+}$ are $4f^{10}$ and $4f^{11}$. The blue dots in the figures indicate the contribution from $4f$ orbitals, and there are several bands near Fermi surface are constituted by $4f$ orbitals. There are also one band cross the Fermi level with blue dots covered, and that is due to the hybridization of $4f$ orbital and itinerant orbitals. The bands mainly constituted by $4f$ orbitals in ErRh$_6$Ge$_4$ is more close to Fermi energy compared with in HoRh$_6$Ge$_4$. While in TmRh$_6$Ge$_4$ the situation is different as displayed in Fig. \ref{fig:band} (d). As the number of $4f$ electrons increasing, the bands close to Fermi energy have more contribution of $4f$ orbitals, and the valley near $A$ point is lifted above Fermi level. 	

It is worth mentioning that the topological properties of the band structure. Due to the $C_{3v}$ group symmetry along the $c$ axis, and the space-inversion symmetry is not present in $ReRh_6Ge_4$, there are triply degenerate nodal points in YRh$_6$Ge$_4$, LaRh$_6$Ge$_4$ and LuRh$_6$Ge$_4$ \cite{Guo2018_PRB98-045134}. If without SOC they all have two pairs of three fold degenerate points along the $\Gamma -A$ direction. The three-fold degenerate points are generated by the crossing of a nondegenerate band and a doubly degenerate band. When SOC is included the three-fold degenerate points changes to a pair of triply degenerate nodal points. A triply degenerate nodal point is considered as intermediate state between double degenerate Weyl point and four-fold degenerate Dirac point. The $C_{3v}$ group symmetry is kept in TmRh$_6$Ge$_4$ since the magnetic axis is along the $c$ axis. There are also triply degenerate points in the band structure of TmRh$_6$Ge$_4$ along the $\Gamma -A$ direction. While in CeRh$_6$Ge$_4$, HoRh$_6$Ge$_4$ and ErRh$_6$Ge$_4$ the the $C_{3v}$ group symmetry is broken by the magnetic axis. The triply degenerate points disappear, leaving the Weyl points. 		

	\subsection{Density of states and Fermi surface} \label{subsec3:DOS-Fermi}
\begin{figure}[htbp]
		\centering
		\includegraphics[width=0.68\textwidth]{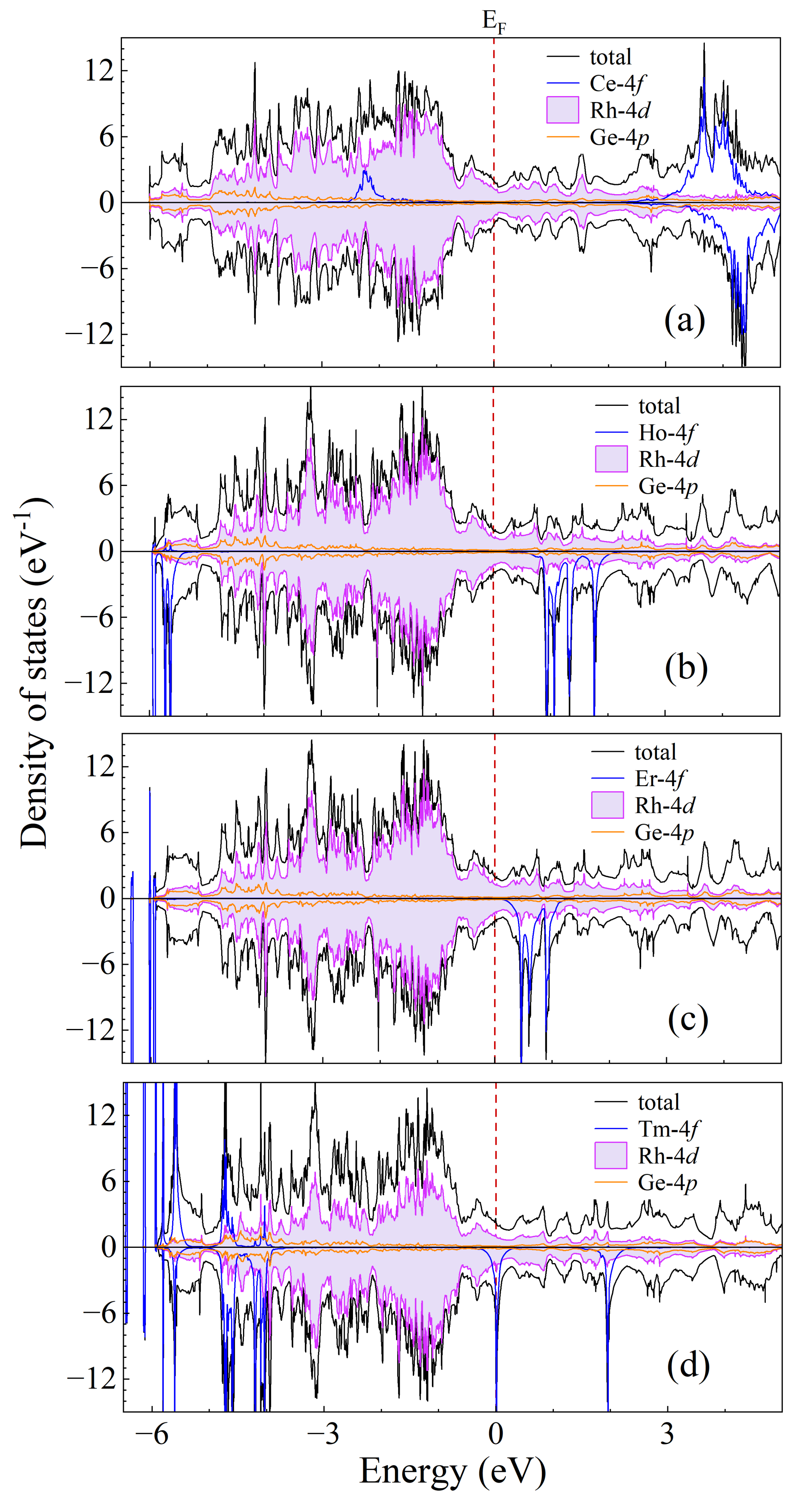} \\
		\caption{\label{fig:DOS} Density of states of (a) CeRh$_6$Ge$_4$, (b) HoRh$_6$Ge$_4$, (c) ErRh$_6$Ge$_4$ and (d) TmRh$_6$Ge$_4$. The calculations are carried in the ferromagnetic phase with the SOC, and by fixing $U=6$ $eV$. The total DOS is displayed with black line,  and the partial DOS of $Ce$-$4f$ orbitals, $Rh$-$4d$ orbitals and $Ge$-$4p$ orbitals are displayed with blue lines, shaped purple lines and orange lines. }		
	\end{figure}

The density of states (DOS) of these compounds are shown in Fig.~\ref{fig:DOS}, and all the calculations are carried in the ferromagnetic case with SOC and $U=6$ $eV$. Above and below the horizontal axis are the majority and minority components of the DOS. The partial DOS of different orbitals are displayed with different colors. Purple shaded areas represent the DOS of $Rh$ $4d$ orbitals, orange lines represent the DOS of $Ge$-4$p$ orbitals, and the blue lines represent the DOS of $4f$ orbitals. The total DOS (black lines in Fig.~\ref{fig:DOS}) is non-zero at Fermi energy, since these compounds are all metallic. The DOS around Fermi energy is mainly contributed by the $4d$ orbitals of $Rh$ ions. As displayed in Fig. \ref{fig:DOS}(a) there is a small blue peak near $-2$ $eV$ below the Fermi energy of CeRh$_6$Ge$_4$. It represents the occupied $4f^1$ state or the $J_{5/2}$ state considering SOC. While in HoRh$_6$Ge$_4$ and TmRh$_6$Ge$_4$ there are a few blue peaks above and close to the Fermi energy. They represent the unoccupied $4f$ orbitals. As more and more $4f$ orbitals are occupied from HoRh$_6$Ge$_4$ to TmRh$_6$Ge$_4$, one peak of the Tm$ $4f$ orbital is located at the Fermi energy. Thus the Tm$ $4f$ orbital is itinerant, and it contributes to the Fermi surface of TmRh$_6$Ge$_4$. As a result TmRh$_6$Ge$_4$ has a large Fermi surface, compared with the other three compounds on which $4f$ orbital is localized and the Fermi surface is small.     
	
\begin{figure}[htbp]
		\centering
		\includegraphics[width=0.68\textwidth]{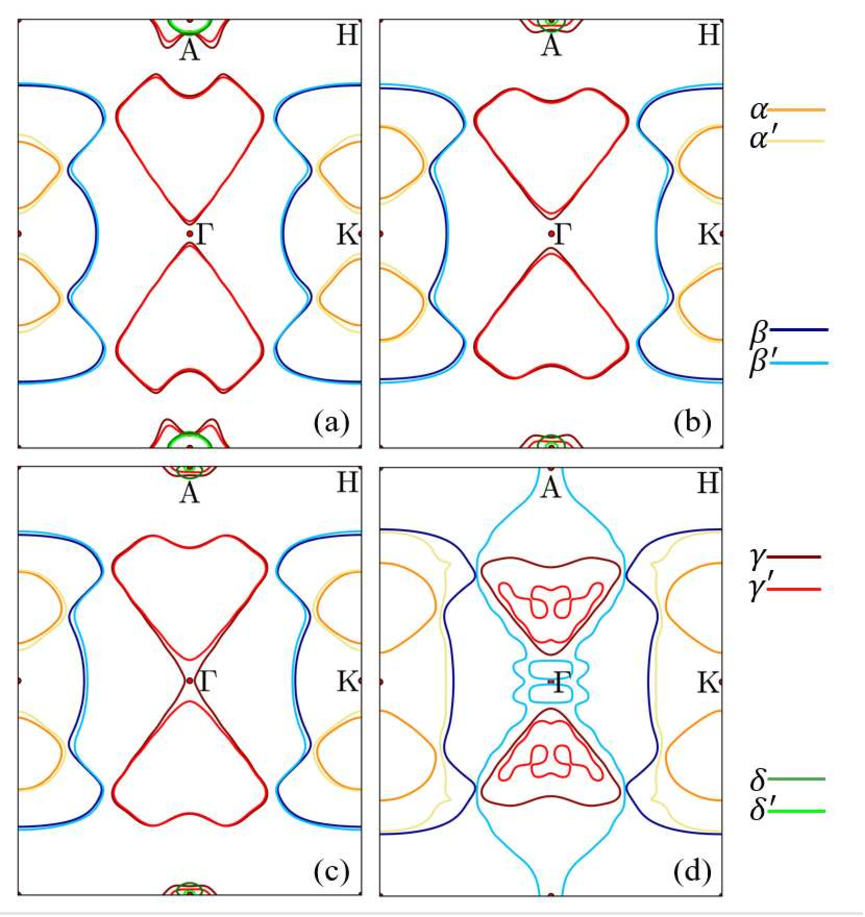} \\
		\caption{\label{fig:2D} The cross session of the $xz$ plane and the Fermi surface  of (a) CeRh$_6$Ge$_4$, (b) HoRh$_6$Ge$_4$, (c) ErRh$_6$Ge$_4$ and (d) TmRh$_6$Ge$_4$ calculated with the SOC and $U=6$ $eV$. The orange line ($\alpha$), yellow line ($\alpha'$), dark blue line ($\beta$) and blue line ($\beta'$) represent the hole type bands, while the brown line ($\gamma$), red line ($\gamma'$), dark green line ($\delta$) and green line ($\delta'$) represent the hole type bands.}		
	\end{figure}

In order to shown the evolution of the Fermi surface volume, we simulate the Fermi surface of these compounds. The CeRh$_6$Ge$_4$ Fermi surface of our simulations are quite similar with previous results \cite{WANG20211389}. There are eight bands cross the Fermi energy, so the Fermi surface of CeRh$_6$Ge$_4$ is composed by eight sheets. Four of them are hole type, and four of them are electron type. The Fermi surface of HoRh$_6$Ge$_4$ and ErRh$_6$Ge$_4$ are composed by eight sheets too, and four of them are electron type. Meanwhile the Fermi surface of TmRh$_6$Ge$_4$ is composed by six sheets, and only two of them are electron type. The cross session of the Fermi surface with the $k_xk_z$ plane displayed in Fig. \ref{fig:2D},  clearly shows the evolution of the Fermi surface. In TmRh$_6$Ge$_4$ the $\delta$ and $\delta'$ bands don't cross the Fermi energy, so the dark green and green lines are absent in Fig. \ref{fig:2D} (d).            
  
	\section{SUMMARY AND OUTLOOK}\label{sec4:Summary}
Based on the empirical experimental evidence that the gemanides compounds $ReRh_6Ge_4$ ($Re=Ce, Ho, Er, Tm$) usually exhibit a ferromagnetic ground state, we have conducted the ferromagnetic DFT calculations on these compounds. By fixing the Coulomb interaction $U=6$ $eV$ and incorporating SOC effects, we predict the magnetic easy axis of CeRh$_6$Ge$_4$ is within $ab$ plane, with an angle of $\varphi=30^{\circ}$ with respect to the $a$ axis. For the less-explored compounds, our predictions reveal that HoRh$_6$Ge$_4$ and ErRh$_6$Ge$_4$ suggest their easy axis along the $ac$ plane, with an angle $\theta=15^{\circ}$ from the $c$ axis. In contrast, TmRh$_6$Ge$_4$ possesses a magnetic easy axis aligned along the $c$ axis, which preserves the $C_{3v}$ point group symmetry. Intriguingly, within the $\Gamma-A$ direction, we observe the presence of triply degenerate nodal points in their band structures. The magnetic axes of the other three compounds, as a result of their unique magnetic properties, break the $C_{3v}$ symmetry, even though the nodal points are absent. Instead, Weyl points emerge along the $\Gamma-A$ direction. The density of states (DOS) analysis further elucidates the almost localized 4$f$ states, and enhancing itinerancy of 4$f$ states from HoRh$_6$Ge$_4$ to TmRh$_6$Ge$_4$. TmRh$_6$Ge$_4$ boasts a substantial Fermi surface due to the active participation of the 4$f$ electrons in its Fermi surface formation.

Recent thermoelectric measurements \cite{Thomas2024_PRB109-L121105} reported an intriguing phenomenon in CeRh$_6$Ge$_4$, where the onset of orbital-selective hybridization within the ferromagnetic phase at around $0.7$ $GPa$ leads to a discernible alteration in the Fermi surface geometry. Despite this, quantum oscillation data indicate that the extremal orbits on the Fermi surface remain unaltered across the critical temperature $T_c$. This discrepancy underscores the need for further numerical investigations on CeRh$_6$Ge$_4$ under pressure to delve deeper into the physics near the quantum critical point (QCP). Meanwhile, experimental studies on the isostructural counterparts, HoRh$_6$Ge$_4$, ErRh$_6$Ge$_4$ and TmRh$_6$Ge$_4$, remain relatively sparse. Our results shall establish a connection between magnetism and electronic structure, facilitating subsequent theoretical and experimental research. 

\section*{acknowledgments}
This work is supported by the  National Key Research and Development Program of China (under Grant No. 2024YFA1408600), the National Natural Science Foundation of China (under Grant No. 12474241), Presidential Foundation of CAEP (under Grant No. YZJJZQ2024014), National Key Research and Development Program of China (under Grant No. 2022YFA1402201).
We are grateful for the fruitful conversations with Pengjie Guo, and we thank Kai Liu for helpful discussions. We acknowledge National Super Computer Center in Tianjin for computing time.

\bibliographystyle{elsarticle-num} 
\bibliography{RE_elsarticle}

@ARTICLE{Belitz1999_PRL82-4707--4710,
  author = {Belitz, D. and Kirkpatrick, T. R. and Vojta, Thomas},
  title = {First Order Transitions and Multicritical Points in Weak Itinerant
	Ferromagnets},
  journal = {Phys. Rev. Lett.},
  year = {1999},
  volume = {82},
  pages = {4707--4710},
  month = {Jun},
  doi = {10.1103/PhysRevLett.82.4707},
  issue = {23},
  numpages = {0},
  publisher = {American Physical Society}
}

@BOOK{wien2k,
  title = {{WIEN2k, An Augmented Plane Wave + Local Orbitals Program for Calculating
	Crystal Properties}},
  publisher = {Karlheinz Schwarz, Techn. Universität Wien, Austria},
  year = {2001},
  author = {P. Blaha and K. Schwarz and G. Madsen and D. Kvasnicka and J. Luitz}
}

@ARTICLE{Brando2016_RMP88-025006,
  author = {Brando, M. and Belitz, D. and Grosche, F. M. and Kirkpatrick, T.
	R.},
  title = {Metallic quantum ferromagnets},
  journal = {Rev. Mod. Phys.},
  year = {2016},
  volume = {88},
  pages = {025006},
  month = {May},
  doi = {10.1103/RevModPhys.88.025006},
  issue = {2},
  numpages = {71},
  publisher = {American Physical Society}
}

@ARTICLE{buschmann1991darstellung,
  author = {Buschmann, Rolf and Schuster, Hans-Uwe},
  title = {Synthesis and Crystal Structure of the Compound {LiCo6P4}},
  journal = {Z. Naturforsch. B},
  year = {1991},
  volume = {46},
  pages = {699--702},
  number = {5},
  doi = {doi:10.1515/znb-1991-0525},
  publisher = {De Gruyter}
}

@ARTICLE{Chen2022_PRB106-075114,
  author = {Chen, Jialin and Wang, Jiangfan and Hu, Danqing and Yang, Yi-feng},
  title = {Continuous ferromagnetic quantum phase transition on an anisotropic
	Kondo lattice},
  journal = {Phys. Rev. B},
  year = {2022},
  volume = {106},
  pages = {075114},
  month = {Aug},
  doi = {10.1103/PhysRevB.106.075114},
  issue = {7},
  numpages = {6},
  publisher = {American Physical Society}
}

@ARTICLE{Chubukov2004_PRL92-147003,
  author = {Chubukov, Andrey V. and P\'epin, Catherine and Rech, Jerome},
  title = {Instability of the Quantum-Critical Point of Itinerant Ferromagnets},
  journal = {Phys. Rev. Lett.},
  year = {2004},
  volume = {92},
  pages = {147003},
  month = {Apr},
  doi = {10.1103/PhysRevLett.92.147003},
  issue = {14},
  numpages = {4},
  publisher = {American Physical Society}
}

@ARTICLE{Custers2003_Nature424-524--527,
  author = {Custers, J. and Gegenwart, P. and Wilhelm, H. and Neumaier, K. and
	Tokiwa, Y. and Trovarelli, O. and Geibel, C. and Steglich, F. and
	Pépin, C. and Coleman, P.},
  title = {The break-up of heavy electrons at a quantum critical point},
  journal = {Nature},
  year = {2003},
  volume = {424},
  pages = {524--527},
  number = {6948},
  issn = {1476-4687},
  refid = {Custers2003}
}

@ARTICLE{Daou_2008,
  author = {R. Daou and Nicolas Doiron-Leyraud and David LeBoeuf and S. Y. Li
	and Francis Lalibert{\'{e}} and Olivier Cyr-Choini{\`{e}}re and Y.
	J. Jo and L. Balicas and J.-Q. Yan and J.-S. Zhou and J. B. Goodenough
	and Louis Taillefer},
  title = {Linear temperature dependence of resistivity and change in the Fermi
	surface at the pseudogap critical point of a high {Tc} superconductor},
  journal = {Nature Physics},
  year = {2008},
  volume = {5},
  pages = {31--34},
  number = {1},
  month = {nov},
  doi = {10.1038/nphys1109},
  publisher = {Springer Science and Business Media {LLC}}
}

@ARTICLE{dorado2013advances,
  author = {B Dorado and M Freyss and B Amadon and M Bertolus and G Jomard and
	P Garcia},
  title = {Advances in first-principles modelling of point defects in {UO2}:
	f electron correlations and the issue of local energy minima},
  journal = {Journal of Physics: Condensed Matter},
  year = {2013},
  volume = {25},
  pages = {333201},
  number = {33},
  month = {jul},
  doi = {10.1088/0953-8984/25/33/333201},
  publisher = {IOP Publishing}
}

@ARTICLE{Guo2018_PRB98-045134,
  author = {Guo, Peng-Jie and Yang, Huan-Cheng and Liu, Kai and Lu, Zhong-Yi},
  title = {Triply degenerate nodal points in $R{\mathrm{Rh}}_{6}{\mathrm{Ge}}_{4}\phantom{\rule{4pt}{0ex}}(R=\text{Y},\text{La},\text{Lu})$},
  journal = {Phys. Rev. B},
  year = {2018},
  volume = {98},
  pages = {045134},
  month = {Jul},
  doi = {10.1103/PhysRevB.98.045134},
  issue = {4},
  numpages = {7},
  publisher = {American Physical Society}
}

@ARTICLE{Hertz1976_PRB14-1165--1184,
  author = {Hertz, John A.},
  title = {Quantum critical phenomena},
  journal = {Phys. Rev. B},
  year = {1976},
  volume = {14},
  pages = {1165--1184},
  month = {Aug},
  doi = {10.1103/PhysRevB.14.1165},
  issue = {3},
  numpages = {0},
  publisher = {American Physical Society}
}

@ARTICLE{Jiao2020_Nature579-523--527,
  author = {Jiao, Lin and Howard, Sean and Ran, Sheng and Wang, Zhenyu and Rodriguez,
	Jorge Olivares and Sigrist, Manfred and Wang, Ziqiang and Butch,
	Nicholas P. and Madhavan, Vidya},
  title = {Chiral superconductivity in heavy-fermion metal UTe2},
  journal = {Nature},
  year = {2020},
  volume = {579},
  pages = {523--527},
  number = {7800},
  __markedentry = {[qiaoni:]},
  abstract = {Spin-triplet superconductors are condensates of electron pairs with
	spin 1 and an odd-parity wavefunction1. An interesting manifestation
	of triplet pairing is the chiral p-wave state, which is topologically
	non-trivial and provides a natural platform for realizing Majorana
	edge modes2,3. However, triplet pairing is rare in solid-state systems
	and has not been unambiguously identified in any bulk compound so
	far. Given that pairing is usually mediated by ferromagnetic spin
	fluctuations, uranium-based heavy-fermion systems containing f-electron
	elements, which can harbour both strong correlations and magnetism,
	are considered ideal candidates for realizing spin-triplet superconductivity4.
	Here we present scanning tunnelling microscopy studies of the recently
	discovered heavy-fermion superconductor UTe2, which has a superconducting
	transition temperature of 1.6 kelvin5. We find signatures of coexisting
	Kondo effect and superconductivity that show competing spatial modulations
	within one unit cell. Scanning tunnelling spectroscopy at step edges
	reveals signatures of chiral in-gap states, which have been predicted
	to exist at the boundaries of topological superconductors. Combined
	with existing data that indicate triplet pairing in UTe2, the presence
	of chiral states suggests that UTe2 is a strong candidate for chiral-triplet
	topological superconductivity.},
  file = {Jiao2020_Nature579-523--527.pdf:2012/Jiao2020_Nature579-523--527.pdf:PDF},
  issn = {1476-4687},
  owner = {qiaoni},
  refid = {Jiao2020},
  timestamp = {2021.07.01}
}

@ARTICLE{Kirkpatrick2020_PRL124-147201,
  author = {Kirkpatrick, T. R. and Belitz, D.},
  title = {Ferromagnetic Quantum Critical Point in Noncentrosymmetric Systems},
  journal = {Phys. Rev. Lett.},
  year = {2020},
  volume = {124},
  pages = {147201},
  month = {Apr},
  doi = {10.1103/PhysRevLett.124.147201},
  issue = {14},
  numpages = {5},
  publisher = {American Physical Society}
}

@ARTICLE{Komijani2018_PRL120-157206,
  author = {Komijani, Yashar and Coleman, Piers},
  title = {Model for a Ferromagnetic Quantum Critical Point in a 1D Kondo Lattice},
  journal = {Phys. Rev. Lett.},
  year = {2018},
  volume = {120},
  pages = {157206},
  month = {Apr},
  doi = {10.1103/PhysRevLett.120.157206},
  issue = {15},
  numpages = {5},
  publisher = {American Physical Society}
}

@ARTICLE{Kotegawa2019_JPSJ88-093702,
  author = {Kotegawa ,Hisashi and Matsuoka ,Eiichi and Uga ,Toshiaki and Takemura
	,Masaki and Manago ,Masahiro and Chikuchi ,Noriyasu and Sugawara
	,Hitoshi and Tou ,Hideki and Harima ,Hisatomo},
  title = {Indication of Ferromagnetic Quantum Critical Point in Kondo Lattice
	CeRh6Ge4},
  journal = {J. Phys. Soc. Jpn.},
  year = {2019},
  volume = {88},
  pages = {093702},
  number = {9},
  doi = {10.7566/JPSJ.88.093702}

}

@ARTICLE{PhysRevLett.72.3262,
  author = {L\"ohneysen, H. v. and Pietrus, T. and Portisch, G. and Schlager,
	H. G. and Schr\"oder, A. and Sieck, M. and Trappmann, T.},
  title = {Non-Fermi-liquid behavior in a heavy-fermion alloy at a magnetic
	instability},
  journal = {Phys. Rev. Lett.},
  year = {1994},
  volume = {72},
  pages = {3262--3265},
  month = {May},
  doi = {10.1103/PhysRevLett.72.3262},
  issue = {20},
  numpages = {0},
  publisher = {American Physical Society}
}

@ARTICLE{legros:cea-02086419,
  author = {Legros, A. and Benhabib, S. and Tabis, Wojciech and Lalibert{\'e},
	F. and Dion, M. and Lizaire, M. and Vignolle, Baptiste and Vignolles,
	David and Raffy, H. and Li, Z. and Auban-Senzier, P. and Doiron-Leyraud,
	N. and Fournier, P. and Colson, Doroth{\'e}e and Taillefer, L. and
	Proust, Cyril},
  title = {{Universal $T$-linear resistivity and Planckian dissipation in overdoped
	cuprates}},
  journal = {{Nature Physics}},
  year = {2019},
  volume = {15},
  pages = {142-147},
  number = {2},
  month = Feb,
  doi = {10.1038/s41567-018-0334-2},
  hal_id = {cea-02086419},
  hal_version = {v1},
  publisher = {{Nature Publishing Group}}
}

@ARTICLE{PhysRevB.91.035130,
  author = {Lengyel, E. and Macovei, M. E. and Jesche, A. and Krellner, C. and
	Geibel, C. and Nicklas, M.},
  title = {Avoided ferromagnetic quantum critical point in {CeRuPO}},
  journal = {Phys. Rev. B},
  year = {2015},
  volume = {91},
  pages = {035130},
  month = {Jan},
  doi = {10.1103/PhysRevB.91.035130},
  issue = {3},
  numpages = {11},
  publisher = {American Physical Society}
}

@ARTICLE{jiaolin:586,
  author = {JIAO Lin},
  title = {Heavy fermion superconductors{[J]}},
  journal = {Physics},
  year = {2020},
  volume = {49},
  pages = {586-594},
  number = {9},
  eid = {586},
  doi = {10.7693/wl20200903},
  numpages = {8},
  publisher = {Physics}
}

@ARTICLE{Liu2023_PNAS120-e2300903120,
  author = {Chia-Chuan Liu and Silke Paschen and Qimiao Si},
  title = {Quantum criticality enabled by intertwined degrees of freedom},
  journal = {Proc. Nat. Acad. Sci.},
  year = {2023},
  volume = {120},
  pages = {e2300903120},
  number = {30},
  doi = {10.1073/pnas.2300903120}
}

@ARTICLE{LiYu2021_ActaPhys.Sin.70-106-136,
  author = {Li Yu, Sheng Yu-Tao, Yang Yi-Feng},
  title = {Research progress of heavy fermion superconductivity theory and materials},
  journal = {Acta Phys. Sin.},
  year = {2021},
  volume = {70},
  pages = {106-136},
  doi = {10.7498/aps.70.20201418},
  issn = {1000-3290},
  printed = {printed}
}

@ARTICLE{Luo2023_PRB108-195146,
  author = {Luo, Shuaishuai and Du, Feng and Su, Dajun and Zhang, Yongjun and
	Zhang, Jiawen and Xu, Jiacheng and Chen, Yuxin and Cao, Chao and
	Smidman, Michael and Steglich, Frank and Yuan, Huiqiu},
  title = {Direction-dependent switching of carrier type enabled by Fermi surface
	geometry},
  journal = {Phys. Rev. B},
  year = {2023},
  volume = {108},
  pages = {195146},
  month = {Nov},
  doi = {10.1103/PhysRevB.108.195146},
  issue = {19},
  numpages = {7},
  publisher = {American Physical Society}
}

@ARTICLE{Millis1993_PRB48-7183--7196,
  author = {Millis, A. J.},
  title = {Effect of a nonzero temperature on quantum critical points in itinerant
	fermion systems},
  journal = {Phys. Rev. B},
  year = {1993},
  volume = {48},
  pages = {7183--7196},
  month = {Sep},
  doi = {10.1103/PhysRevB.48.7183},
  issue = {10},
  numpages = {0},
  publisher = {American Physical Society}
}

@BOOK{Moriya1985,
  title = {Spin fluctuations in Itinerant Electron Magnetism},
  publisher = {Springer, Berlin},
  year = {1985},
  author = {T. Moriya}
}

@ARTICLE{murnaghan1944compressibility,
  author = {Murnaghan, Francis Dominic},
  title = {The compressibility of media under extreme pressures},
  journal = {Proc. Nat. Acad. Sci.},
  year = {1944},
  volume = {30},
  pages = {244--247},
  number = {9},
  doi = {10.1073/pnas.30.9.244},
  publisher = {National Acad Sciences}
}

@ARTICLE{Pei2021_PRB103-L180409,
  author = {Pei, Y. H. and Zhang, Y. J. and Wei, Z. X. and Chen, Y. X. and Hu,
	K. and Yang, Yi-feng and Yuan, H. Q. and Qi, J.},
  title = {Unveiling the hybridization process in a quantum critical ferromagnet
	by ultrafast optical spectroscopy},
  journal = {Phys. Rev. B},
  year = {2021},
  volume = {103},
  pages = {L180409},
  month = {May},
  doi = {10.1103/PhysRevB.103.L180409},
  issue = {18},
  numpages = {6},
  publisher = {American Physical Society}
}

@ARTICLE{PhysRevLett.77.3865,
  author = {Perdew, John P. and Burke, Kieron and Ernzerhof, Matthias},
  title = {Generalized Gradient Approximation Made Simple},
  journal = {Phys. Rev. Lett.},
  year = {1996},
  volume = {77},
  pages = {3865--3868},
  month = {Oct},
  doi = {10.1103/PhysRevLett.77.3865},
  issue = {18},
  numpages = {0},
  publisher = {American Physical Society}
}

@ARTICLE{Ran2019_Science365-684,
  author = {Sheng Ran and Chris Eckberg and Qing-Ping Ding and Yuji Furukawa
	and Tristin Metz and Shanta R. Saha and I-Lin Liu and Mark Zic and
	Hyunsoo Kim and Johnpierre Paglione and Nicholas P. Butch},
  title = {Nearly ferromagnetic spin-triplet superconductivity},
  journal = {Science},
  year = {2019},
  volume = {365},
  pages = {684},
  doi = {10.1126/science.aav8645}
}

@ARTICLE{Shen_2020,
  author = {Bin Shen and Yongjun Zhang and Yashar Komijani and Michael Nicklas
	and Robert Borth and An Wang and Ye Chen and Zhiyong Nie and Rui
	Li and Xin Lu and Hanoh Lee and Michael Smidman and Frank Steglich
	and Piers Coleman and Huiqiu Yuan},
  title = {Strange-metal behaviour in a pure ferromagnetic Kondo lattice},
  journal = {Nature},
  year = {2020},
  volume = {579},
  pages = {51--55},
  number = {7797},
  month = {mar},
  doi = {10.1038/s41586-020-2052-z},
  publisher = {Springer Science and Business Media {LLC}}
}

@ARTICLE{Shu2021_PRB104-L140411,
  author = {Shu, J. W. and Adroja, D. T. and Hillier, A. D. and Zhang, Y. J.
	and Chen, Y. X. and Shen, B. and Orlandi, F. and Walker, H. C. and
	Liu, Y. and Cao, C. and Steglich, F. and Yuan, H. Q. and Smidman,
	M.},
  title = {Magnetic order and crystalline electric field excitations of the
	quantum critical heavy-fermion ferromagnet $\mathrm{Ce}{\mathrm{Rh}}_{6}{\mathrm{Ge}}_{4}$},
  journal = {Phys. Rev. B},
  year = {2021},
  volume = {104},
  pages = {L140411},
  month = {Oct},
  doi = {10.1103/PhysRevB.104.L140411},
  issue = {14},
  numpages = {6},
  publisher = {American Physical Society}
}

@ARTICLE{Si2010_Science329-1161--1166,
  author = {Si, Qimiao and Steglich, Frank},
  title = {Heavy Fermions and Quantum Phase Transitions},
  journal = {Science},
  year = {2010},
  volume = {329},
  pages = {1161},
  number = {5996},
  doi = {10.1126/science.1191195},
  issn = {0036-8075},
  publisher = {American Association for the Advancement of Science}
}

@ARTICLE{RevModPhys.73.797,
  author = {Stewart, G. R.},
  title = {Non-Fermi-liquid behavior in $d$- and $f$-electron metals},
  journal = {Rev. Mod. Phys.},
  year = {2001},
  volume = {73},
  pages = {797--855},
  month = {Oct},
  doi = {10.1103/RevModPhys.73.797},
  issue = {4},
  numpages = {0},
  publisher = {American Physical Society}
}

@ARTICLE{PhysRevLett.105.217201,
  author = {Taufour, V. and Aoki, D. and Knebel, G. and Flouquet, J.},
  title = {Tricritical Point and Wing Structure in the Itinerant Ferromagnet
	{UGe2}},
  journal = {Phys. Rev. Lett.},
  year = {2010},
  volume = {105},
  pages = {217201},
  month = {Nov},
  doi = {10.1103/PhysRevLett.105.217201},
  issue = {21},
  numpages = {4},
  publisher = {American Physical Society}
}

@ARTICLE{Thomas2024_PRB109-L121105,
  author = {Thomas, S. M. and Seo, S. and Asaba, T. and Ronning, F. and Rosa,
	P. F. S. and Bauer, E. D. and Thompson, J. D.},
  title = {Probing quantum criticality in ferromagnetic ${\mathrm{CeRh}}_{6}{\mathrm{Ge}}_{4}$},
  journal = {Phys. Rev. B},
  year = {2024},
  volume = {109},
  pages = {L121105},
  month = {Mar},
  doi = {10.1103/PhysRevB.109.L121105},
  issue = {12},
  numpages = {6},
  publisher = {American Physical Society}
}

@ARTICLE{PhysRevLett.85.626,
  author = {Trovarelli, O. and Geibel, C. and Mederle, S. and Langhammer, C.
	and Grosche, F. M. and Gegenwart, P. and Lang, M. and Sparn, G. and
	Steglich, F.},
  title = {{YbRh2Si2}:Pronounced Non-Fermi-Liquid Effects above a Low-Lying
	Magnetic Phase Transition},
  journal = {Phys. Rev. Lett.},
  year = {2000},
  volume = {85},
  pages = {626--629},
  month = {Jul},
  doi = {10.1103/PhysRevLett.85.626},
  issue = {3},
  numpages = {0},
  publisher = {American Physical Society}
}

@ARTICLE{PhysRevLett.93.256404,
  author = {Uhlarz, M. and Pfleiderer, C. and Hayden, S. M.},
  title = {Quantum Phase Transitions in the Itinerant Ferromagnet {ZrZn2}},
  journal = {Phys. Rev. Lett.},
  year = {2004},
  volume = {93},
  pages = {256404},
  month = {Dec},
  doi = {10.1103/PhysRevLett.93.256404},
  issue = {25},
  numpages = {4},
  publisher = {American Physical Society}
}

@ARTICLE{Vosswinkel2013_Zfauac639-2623,
  author = {Voßwinkel, Daniel and Niehaus, Oliver and Pöttgen, Rainer},
  title = {New Rhodium-rich Germanides {RERh6Ge4 (RE = Y, La, Pr, Nd, Sm–Lu)}},
  journal = {Z. Anorg. Allg. Chem.},
  year = {2013},
  volume = {639},
  pages = {2623-2630},
  number = {14},
  doi = {https://doi.org/10.1002/zaac.201300369}
}

@ARTICLE{Vosswinkel2012_ZfNatB67-1241,
  author = {Daniel Voßwinkel and Oliver Niehaus and Ute Ch. Rodewald and Rainer
	Pöttgen},
  title = {Bismuth Flux Growth of CeRh6Ge4 and CeRh2Ge2 Single Crystals},
  journal = {Z. Naturforsch. B},
  year = {2012},
  volume = {67},
  pages = {1241--1247},
  number = {12},
  doi = {doi:10.5560/znb.2012-0265}
}

@ARTICLE{WANG20211389,
  author = {An Wang and Feng Du and Yongjun Zhang and David Graf and Bin Shen
	and Ye Chen and Yang Liu and Michael Smidman and Chao Cao and Frank
	Steglich and Huiqiu Yuan},
  title = {Localized 4f-electrons in the quantum critical heavy fermion ferromagnet
	{CeRh6Ge4}},
  journal = {Science Bulletin},
  year = {2021},
  volume = {66},
  pages = {1389-1394},
  number = {14},
  doi = {https://doi.org/10.1016/j.scib.2021.03.006},
  issn = {2095-9273}
}

@ARTICLE{Wang2022_SCPMA65-257211,
  author = {Wang, Jiangfan and Yang, Yi-Feng},
  title = {A unified theory of ferromagnetic quantum phase transitions in heavy
	fermion metals},
  journal = {Sci. China-Phys. Mech. Astron.},
  year = {2022},
  volume = {65},
  pages = {257211},
  number = {5},
  issn = {1869-1927},
  refid = {Wang2022}
}

@ARTICLE{Wu_2021,
  author = {Yi Wu and Yongjun Zhang and Feng Du and Bin Shen and Hao Zheng and
	Yuan Fang and Michael Smidman and Chao Cao and Frank Steglich and
	Huiqiu Yuan and Jonathan D. Denlinger and Yang Liu},
  title = {Anisotropic $c\ensuremath{-}f$ Hybridization in the Ferromagnetic
	Quantum Critical Metal {CeRh6Ge4}},
  journal = {Phys. Rev. Lett.},
  year = {2021},
  volume = {126},
  pages = {216406},
  month = {May},
  doi = {10.1103/PhysRevLett.126.216406},
  issue = {21},
  numpages = {6},
  publisher = {American Physical Society}
}

@ARTICLE{Xu2021_CPL38-087101,
  author = {Jia-Cheng Xu and Hang Su and Rohit Kumar and Shuai-Shuai Luo and
	Zhi-Yong Nie and An Wang and Feng Du and Rui Li and Michael Smidman
	and Hui-Qiu Yuan},
  title = {Ce-Site Dilution in the Ferromagnetic Kondo Lattice CeRh$_6$Ge$_4$},
  journal = {Chin. Phys. Lett.},
  year = {2021},
  volume = {38},
  pages = {087101},
  number = {8},
  eid = {087101},
  doi = {10.1088/0256-307X/38/8/087101},
  publisher = {Chin. Phys. Lett.}
}

@ARTICLE{Zhang2022_PRB106-054409,
  author = {Zhang, Y. J. and Nie, Z. Y. and Li, R. and Li, Y. C. and Yang, D.
	L. and Shen, B. and Chen, Y. and Du, F. and Luo, S. S. and Su, H.
	and Shi, R. and Wang, S. Y. and Nicklas, M. and Steglich, F. and
	Smidman, M. and Yuan, H. Q.},
  title = {Suppression of ferromagnetism and influence of disorder in silicon-substituted
	${\mathrm{CeRh}}_{6}{\mathrm{Ge}}_{4}$},
  journal = {Phys. Rev. B},
  year = {2022},
  volume = {106},
  pages = {054409},
  month = {Aug},
  doi = {10.1103/PhysRevB.106.054409},
  issue = {5},
  numpages = {7},
  publisher = {American Physical Society}
}






\end{document}